\documentclass[aoas,MSNbibl,nameyear,dvips]{arximspdf}
\usepackage{dcolumn,multirow}
\usepackage{graphicx}

\doi{10.1214/14-AOAS725} %kopijuoti is PTS
\volume{8}
\issue{2}
\pubyear{2014}
\firstpage{726}
\lastpage{746}

\makeatletter
\setattribute{frontmatter}{skip}{0\p@ plus 3\p@ minus 3\p@}
\setattribute{frontmatter}{cmd}{}
\setattribute{abstract}{width}{280pt}
\newcolumntype{d}[1]{D{.}{.}{#1}}
\renewcommand{\mid}{|}
\newcommand{\ta}{.}
\newproclaim{assumption}{Assumption}
\makeatother

\begin{document}
\begin{frontmatter}

\title{Maximum likelihood and pseudo score approaches for parametric
time-to-event analysis with informative entry times\thanksref{T1}}
\runtitle{Time-to-event analysis with informative entry times}

\begin{aug}
\author{\fnms{Brian D. M.} \snm{Tom}\corref{}\ead[label=e1]{brian.tom@mrc-bsu.cam.ac.uk}},
\author{\fnms{Vernon T.} \snm{Farewell}\ead[label=e2]{vern.farewell@mrc-bsu.cam.ac.uk}}
\and
\author{\fnms{Sheila M.} \snm{Bird}\ead[label=e3]{sheila.brid@mrc-bsu.cam.ac.uk}}
\runauthor{B. D. M. Tom, V. T. Farewell and S. M. Bird}
\affiliation{MRC Biostatistics Unit}
\address{MRC Biostatistics Unit\\
Robinson Way\\
Cambridge CB2 0SR\\
United Kingdom\\
\printead{e1}\\
\phantom{E-mail:\ }\printead*{e2}\\
\phantom{E-mail:\ }\printead*{e3}}%adresu isvedimo komanda gale!
\end{aug}
\thankstext{T1}{Supported in part by the Medical Research Council
(Unit Programme numbers U105261167, U105260794).}

% HISTORY:
\received{\smonth{3} \syear{2013}}
\revised{\smonth{9} \syear{2013}}

% ABSTRACT
%
\begin{abstract}
We develop a maximum likelihood estimating approach for time-to-event Weibull
regression models with outcome-dependent sampling, where sampling of
subjects is
dependent on the residual fraction of the time left to developing the
event of interest.
Additionally, we propose a two-stage approach which proceeds by iteratively
estimating, through a pseudo score, the Weibull parameters of interest
(i.e., the regression parameters)
conditional on the inverse probability of sampling weights; and then
re-estimating these weights
(given the updated Weibull parameter estimates) through the profiled
full likelihood.
With these two new methods, both the estimated sampling mechanism parameters
and the Weibull parameters are consistently estimated under correct
specification of the
conditional referral distribution. Standard errors for the regression
parameters are obtained
directly from inverting the observed information matrix in the full
likelihood specification and
by either calculating bootstrap or robust standard errors for the
hybrid pseudo score/profiled
likelihood approach. Loss of efficiency with the latter approach is
considered. Robustness of the
proposed methods to misspecification of the referral mechanism and the
time-to-event distribution
is also briefly examined. Further, we show how to extend our methods to
the family of parametric
time-to-event distributions characterized by the generalized gamma
distribution. The motivation for
these two approaches came from data on time to cirrhosis from hepatitis
C viral infection in patients
referred to the Edinburgh liver clinic. We analyze these data here.
\end{abstract}

% KEYWORDS
% Pirmas kwd is didziosios raides
%

\end{frontmatter}

%s1 #&#
\section{Introduction}\label{sec1}

The modeling of the time from disease onset or infection (i.e.,
initiating event) to an outcome of relevance is of considerable
importance\vadjust{\goodbreak} in studies of the natural history of a disease and in
projection of disease burden. Prospective studies which recruit and
follow an appropriate cohort of subjects from disease onset to the
event of interest are ideal for this purpose. However, these studies
are inefficient in terms of resources if the event of interest tends
to occur well after disease onset, as is the case for hepatitis C virus
(HCV) studies of progression to cirrhosis from initial infection. The
alternative is to follow a prevalent cohort of cross-sectionally sampled
subjects who, prior to recruitment, have already experienced the
initiating event (e.g., HCV infection) but not yet the event of
interest (e.g., cirrhosis).
The left truncated time-to-event data obtained from such a study
provide a length-biased sample of the incident population, if sampling
is such
that an assumption of stationarity over calendar time for the
occurrence of
the initiating event can be made. Methods for handling both incidence
data and
such length-biased prevalence data have been well described in the
(bio)statistics literature [\citet{AndersenBorganGill1993},
\citet{WangBrookmeyerJewell1993},
\citet{KalbfleischPrentice2002}, \citet{Brookmeyer2005},
\citet{Keiding2005},
\citet{Wang2005}, \citet{Tsai2009}, \citet{QinShen2010}].

A less explored situation is the analysis of prevalence data arising
from a referral
cohort where entry into the cohort is dependent on a subject's residual
fraction of
time remaining to the event of interest, and inference on the incident
population is required.
Such data are believed to occur in HCV studies conducted in tertiary
care settings, where
HCV patients are more likely to be referred to specialist clinics at
later stages of disease
[\citet{FuTomBird2007}]. The conventional truncation likelihood
approach which simply conditions on
the time of entry into the cohort does not work here, as the referral
time and the time to the event are correlated.
The ignoring of this referral bias has led to higher rates of progression
to cirrhosis being reported in studies in specialist clinics compared
to those in community-based settings
[\citet{FreemanDoreLaw2001}]. As cirrhosis linked to HCV infection
is a major epidemic of the 21st century,
it is extremely important to get an accurate picture of the present and
future disease burden facing
affected regions in order to inform public health decisions and actions.

The aforementioned type of referral or outcome-dependent sampling bias
is particularly difficult to
deal with unless a full specification (up to unknown parameters) of the
probability sampling
generating mechanism is provided. In practice, this mechanism will
rarely be known and, instead,
an approximate formulation of the sampling distribution, which is
reasonably robust to misspecification,
would be sought.

Previously, \citet{FuTomBird2009} proposed a weighted pseudo
score [\citet{Lawless1997}, \citet{CookLawless2007}]
or inverse probability weighted method for estimating the parameters of
a Weibull regression model for the
incubation period from infection to cirrhosis for the community of
hepatitis C virus-infected individuals, when
there is cirrhosis-related referral bias to the studied prevalent
cohort. The method assumed that everyone in
the community would come to clinical attention at or before cirrhosis,
so that cirrhosis events are not missed.
Therefore, the target community population was assumed ``immortal'' (in
the sense of no competing events), and individuals
observed in the study sample to have experienced a cirrhotic event were
associated with a weight of one in the estimation procedure.
However, for other individuals, \citet{FuTomBird2009} used
approximate weights and, therefore, consistency of these estimated
weights, and,
consequently, the regression parameter estimates of interest, was, in
general, not guaranteed.

Here we outline a full likelihood approach to this outcome-dependent referral
problem in which the likelihood for the joint distribution of the time
to referral and
the time to outcome of interest, both from the initiating event, is
fully specified. In
practice, depending on the dimensionality of the joint parameter space,
the full likelihood
may be difficult to maximize over both the regression parameters of
interest and the parameters
associated with the time-to-entry process. Therefore, we also
investigate another strategy
based on a hybrid two-stage approach that iteratively alternates
between estimating
the parameters associated with the time-to-outcome distribution (i.e.,
regression and shape parameters)
from a pseudo score with fixed weights and then estimating the parameters
associated with the time-to-entry/referral process from the profiled
full likelihood
assuming the regression and shape parameters are known. We retain
the assumption of an immortal cohort, although this can be relaxed
[\citet{CopasFarewell2001}].
Primarily, we describe the approaches where the time-to-event
distribution is assumed Weibull.
However, we show how the methods can be extended to the family of
parametric time-to-event
distributions characterized by the generalized gamma distribution
[\citet{Stacy1962}, \citet{StacyMihram1965}, \citet
{Prentice1974},
\citet{FarewellPrentice1977}, \citet{Lawless1980}, \citet
{CoxChu2007}], for which the
Weibull is an important special case.

%s2 #&#
\section{Notation, framework and assumptions}\label{sec2} For individuals in the
target/\break incident population, let the calendar time of the initiating
event be $Y$ and the calendar period of interest for inference on this
population be between calendar times $d_1$~and~$d_2$.
Therefore, $d_1 \leq Y \leq d_2$. Clinical observation of an individual
will be
left truncated at their time of referral to the clinic which is the
time of entry into the cohort for those referred before $d_{2}$. Let
the time intervals from $Y$ to potential referral and to the event of interest
be $R$ and $T$, respectively, and denote by $Z$ the $p \times1$ vector of
explanatory variables. We assume that the time-to-event $T$ from $Y$
in the incident population comes from a Weibull distribution with
support on the positive real line and with positive shape and scale
parameters, $\gamma$ and $\lambda$, respectively, where $\lambda=
\exp(\beta^Tz)$ for given $Z=z$ and $\beta$ is a vector of regression
parameters associated with $z$. More explicitly, the density and
distribution functions of $T$ from an initiating event calendar time
$Y=y$, and given the vector of explanatory variables $Z=z$, are
$f_{T}(t \mid y,z)= \{\gamma\exp(-\gamma\beta^Tz)\} \exp[-
\{t/\exp(\beta^Tz)\}^\gamma] t^{\gamma-1}$ and $F_{T}(t \mid y,z)=1 -
\exp[-\{t/\exp(\beta^Tz)\}^\gamma]$, respectively. As there is no
dependence on the actual value of $y$ in these functions, we simplify
the notation
for the density and distribution functions of $T$ to $f_{T}(t \mid z)$
and $F_{T}(t \mid z)$,
respectively. Additionally, we assume, as is done for length-biased
sampling problems,
that within the calendar period $[d_1,d_2]$, the rate of occurrence of
the initiating event
remains constant. The independence of the distribution of $T$ from when its
initiating event occurred and the stationarity of the initiating event
process within the
calendar period of interest are together referred to as the steady
state or
equilibrium condition [\citet{Wang2005}].

An individual is assumed to be included in the studied prevalent
cohort if $0 < R < d_2 - Y$, with $S = I(0 < R < d_2-Y)$ the indicator
variable denoting selection/inclusion. In addition to the assumption
that selected patients will experience the event of interest and be
referred prior to the time of the event, we assume the following for
the individuals in the target population.

%
%as1 #&#
\begin{assumption}[(Truncation before outcome)]
\label{assumption1}
The truncation (or potential referral or entry) time of an individual
is always less than the time to outcome and
so $R < T$.
\end{assumption}

%
%as2 #&#
\begin{assumption}[(Conditional truncation time)]
\label{assumption2}
For a known vector $\nu= (\nu_0,\ldots,\nu_{m+1})^T$, with $\nu_0=0$,
$\nu_{m+1}=1$ and $\nu_j < \nu_{j+1}$ $(j=0,\ldots,m)$, and\vspace*{1.5pt} unknown
mixture probability vector
$\pi'=(\pi_0,\ldots, \pi_m)^T$ with $\sum_{j=0}^{m}\pi_j=1$, the
distribution of $R$ given $T=t$ (for $t>0$) is a mixture of independent
uniform random variables
with support in the interval $[0,t)$, density function
\begin{eqnarray*}
f_{R \mid T}(r \mid t) &=& \sum_{j=0}^{m}
\frac{\pi_j}{(\nu_{j+1}-\nu_j)t}I(\nu_j < r/t \leq\nu_{j+1})
\end{eqnarray*}
and cumulative distribution function
\begin{eqnarray*}
F_{R \mid T}(r \mid t) &=& \sum_{j=0}^{m}
\frac{\pi_j\{ \min(r,\nu
_{j+1}t) -
\max(0,\nu_j t) \}}{(\nu_{j+1}-\nu_j)t}I(\nu_j < r/t).
\end{eqnarray*}
\end{assumption}
The form chosen for this conditional density reflects the belief that
the residual fraction, $1-r/t$, of time
remaining to the event of interest (or, alternatively, the fraction,
$r/t$, of event time elapsed) drives whether a subject
is referred [\citet{FuTomBird2009}]. It is constructed as a mixture
of uniforms so as to allow flexibility in the shape of distribution
that can be captured. A notable feature of the random variable $V=R/T$
(for $T>0$), corresponding to
the fraction of time elapsed to the event of interest, is its
independence from $T$ (see theorem in the supplementary material [\citet{supp}]).
We will subsequently investigate the impact of misspecifying the
partitioning of $\nu$ on results obtained.

For selected subjects ($S=1$), denote by $C$ the censoring time from
entry into the cohort, and let $X=\min(T,R+C)$
be the observed follow-up\vadjust{\goodbreak} time until the outcome event or censoring,
with $\Delta= I(T-R<C)$ the ``right censoring''
indicator taking the value $1$ when uncensored. As the calendar period
of interest for inference on this population
is between $d_1$ and $d_2$, then for selected subjects, $d_2-Y \geq X$.
That is, follow-up beyond $d_2$ is not planned.
Additionally, we assume that $(T,R)$ is independent of $C$ [conditional
on either $Z$ or $(Z,Y)$] and that the parameters
governing the distribution of $C$ are distinct from those governing the
joint distribution of $(T,R)$. That is,
the censoring process is ignorable.

To proceed with estimation, we make the following further simplifying
assumption:

%
%as3 #&#
\begin{assumption}[(Known initiation time)]
\label{assumption3}
The calendar time of the initiating event can be determined for those subjects
selected for inclusion in the cohort.
\end{assumption}

\begin{figure}%[ht]

\includegraphics{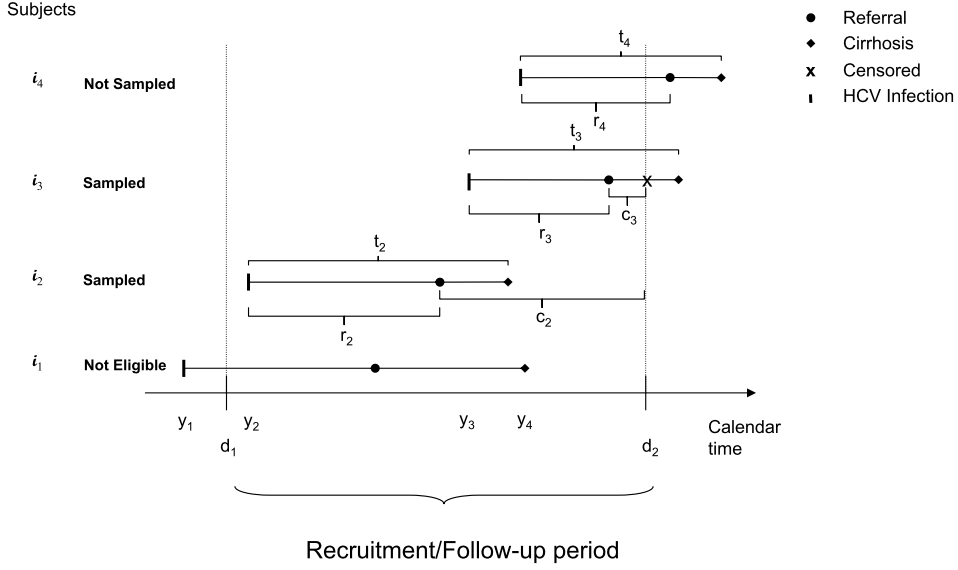}

\caption{Prevalent referral cohort sampling setup.}\label{fig1}
\end{figure}

In Section~\ref{discuss} we discuss how one would proceed if the time
of the initiating event is best known to within an interval.
Figure~\ref{fig1} presents pictorially the salient features of our
prevalent referral cohort design setup.

%
%f1 #&#

%s3 #&#
\section{Estimation methods}\label{sec3}
%s3.1 #&#
\subsection{Maximum likelihood approach}\label{sec3.1}
Let $n$ be the number of individuals who have been selected into the
cohort. For
an included individual $i \in\{1,\ldots,n\}$, let the observed data be
$(r_i,x_i,\delta_i,y_i,z_i)$, which are assumed to be
independent realizations of $(R_i, X_i, \Delta_i, Y_i, Z_i)$. Under
Assumptions~\ref{assumption1} and \ref{assumption3}, the ignorability
of the
censoring process and conditional on $\{Z_i\}$ and $\{Y_i\}$, the full
likelihood for $\theta^T = (\gamma, \beta^T, \pi^T)$, where $\pi
=(\pi_1,\ldots, \pi_m)^T$, can be written (with, for conciseness, some
abuse of
notation for
continuous variables)
as
%
%
%e1 #&#
\begin{eqnarray}
L(\theta)&=& \prod_{i=1}^{n}\bigl\{
\operatorname{pr}(R_i=r_i, T_i=x_i
\mid Y_i=y_i, Z_i=z_i,
S_i=1)^{\delta_i}
\nonumber
\\
&&\hspace*{15pt}{} \times \operatorname{pr}(R_i=r_i, T_i \geq
x_i \mid Y_i=y_i, Z_i=z_i,
S_i=1)^{1-\delta_i}\bigr\} \label{like1}
\\
&=& \prod_{i=1}^nL_i(
\theta).
\nonumber
\end{eqnarray}
The first term in the product is the likelihood contribution if $x_i$
corresponds to the true
time-to-event $t_i$ (i.e., $\delta_i=1$) and the second when a right censored
event time is observed (i.e., $\delta_i=0$).

When $\delta_i=1$ and setting $u_i=d_2-y_i$, it can be shown that
\begin{eqnarray*}
&& \operatorname{pr}(R_i=r_i, T_i=x_i
\mid Y_i=y_i, Z_i=z_i,
S_i=1)
\\
&&\qquad = \frac
{f_{R \mid T}(r_i
\mid x_i) f_{T}(x_i \mid z_i)}{\operatorname{pr}(0 < R_i < u_i)}.
\end{eqnarray*}
In the situation where $\gamma> 1$ (i.e., the hazard rate of $T$
increases over
time), and defining $\varphi= (\gamma- 1)/\gamma$, the denominator,
$\operatorname{pr}(0 < R_i < u_i)$,
can be analytically evaluated and is found to be
\begin{eqnarray}
\label{one} & & \sum_{j=0}^{m}
\frac{\pi_j}{(\nu_{j+1}-\nu_j)} \bigl[ \frac{}{} \bigl\{ \nu_{j+1}
F_{T}(u_i/\nu_{j+1} \mid z_i) -
\nu_j F_{T}(u_i/\nu_j \mid
z_i) \bigr\} \nonumber
\\
&&\hspace*{77pt} {}+u_ie^{-\beta^Tz_i}\Gamma(
\varphi)
\bigl\{F_G\bigl((u_i/\nu_{j})^\gamma;
e^{-\gamma\beta^Tz_i},\varphi\bigr)\nonumber
\\
&&\hspace*{155pt}{}  - F_G\bigl((u_i/
\nu_{j+1})^\gamma; e^{-\gamma\beta^Tz_i},\varphi\bigr) \bigr\} \bigr]
\nonumber\\[-8pt]\\[-8pt]
& &\qquad = \sum_{j=0}^{m}\frac{\pi_j}{(\nu_{j+1}-\nu_j)}
\bigl[ \frac
{}{} \bigl\{ \nu_{j+1} F_{T}(u_i/
\nu_{j+1} \mid z_i) - \nu_j
F_{T}(u_i/\nu_j \mid z_i) \bigr
\}\nonumber
\\
&&\hspace*{110pt}{} + u_ie^{-\beta^Tz_i} \bigl\{\Gamma\bigl(\varphi,e^{-
\gamma\beta^Tz_i}(u_i/
\nu_{j+1})^\gamma\bigr)\nonumber
\\
&&\hspace*{173pt}{} - \Gamma\bigl(\varphi,e^{-
\gamma\beta^Tz_i}(u_i/
\nu_{j})^\gamma\bigr) \bigr\} \bigr]
\nonumber
\end{eqnarray}
with $F_G(u; r, s) = \gamma(s,ru)/\Gamma(s) = \{\Gamma(s)
- \Gamma(s,ru)\}/\Gamma(s)$
the cumulative distribution function of a gamma random variable with
rate $r>0$ and shape \mbox{$s>0$},
evaluated at $u$ ($0 < u < \infty$), where $\gamma(s,u)= \int
_0^ut^{s-1}e^{-t}\,dt$
and $\Gamma(s,u)=\int_u^\infty t^{s-1}e^{-t}\,dt$ denote the lower and upper
incomplete gamma functions and $\Gamma(s) = \int_0^\infty
t^{s-1}e^{-t}\,dt$ the
ordinary gamma function. Details of the derivation are provided in the
supplementary material [\citet{supp}] for the family of parametric time-to-event distributions
characterized by
the generalized gamma distribution with either monotonically increasing
or arc shaped (upside-down bathtub)
hazards [\citet{Glaser1980}, \citet{CoxChu2007}].

For selected individuals with $\delta_i=0$, the likelihood
contribution in
(\ref{like1}), $\operatorname{pr}(R_i=r_i, T_i \geq x_i \mid Y_i=y_i,
Z_i=z_i, S_i=1)$, can be written as
\[
\frac{\operatorname{pr}(R_i=r_i, T_i \geq x_i \mid Y_i=y_i,
Z_i=z_i)}{\operatorname{pr}(0 <R_i <u_i)},
\]
where it can be shown (see the supplementary material [\citet{supp}])
that when $\gamma> 1$, the numerator, $\operatorname{pr}(R_i=r_i, T_i \geq
x_i \mid Y_i=y_i, Z_i=z_i) =
\operatorname{pr}(R_i=r_i, T_i \geq x_i \mid Z_i=z_i)$, takes the closed form
%
%
%e2 #&#
\begin{eqnarray}\label
{two}
& & \sum_{j=0}^{m} \frac{\pi_j}{(\nu_{j+1}-\nu_j)}
\bigl[ \frac{}{} \Gamma(\varphi) e^{-\beta^Tz_i}I\bigl(
\nu_j < \min(r_i/x_i,\nu_{j+1})
\bigr)
\nonumber
\\
&&\hspace*{76pt}{}\times \bigl\{ F_G\bigl((r_i/\nu_{j})^\gamma;
e^{-\gamma\beta
^Tz_i},\varphi\bigr)
\nonumber
\\
&&\hspace*{93pt}{}- \frac{}{} F_G\bigl(\bigl(r_i/
\min(r_i/x_i,\nu_{j+1})\bigr)^\gamma;
e^{-\gamma\beta^Tz_i},\varphi\bigr) \bigr\} \bigr]
\nonumber\\[-8pt]\\[-8pt]
&&\qquad = \sum_{j=0}^{m} \frac{\pi_j}{(\nu_{j+1}-\nu_j)}
\bigl[ \frac{}{} e^{-\beta^Tz_i}I\bigl(\nu_j <
\min(r_i/x_i,\nu_{j+1})\bigr) \nonumber
\\
&&\hspace*{109pt}{}\times
\bigl\{ \Gamma\bigl(\varphi,e^{-\gamma\beta^Tz_i}\bigl(r_i/
\min(r_i/x_i,\nu_{j+1})\bigr)^\gamma
\bigr)
\nonumber
\\
&&\hspace*{178pt}{} - \Gamma\bigl(\varphi,e^{-\gamma\beta^Tz_i}(r_i/\nu
_{j})^\gamma\bigr) \bigr\} \bigr].
\nonumber
\end{eqnarray}

For the case where $\gamma< 1$ (i.e., the hazard rate of $T$ is
monotonically decreasing over time),
similar closed-form expressions for $\operatorname{pr}(0 < R_i < u_i)$ and
$\operatorname{pr}(R_i=r_i, T_i \geq x_i \mid Z_i=z_i)$
can be obtained but with the upper incomplete gamma function of the form
$\Gamma(\varphi,(u/\lambda)^\gamma)$ replaced with $(u/\lambda
)^{\gamma\varphi}
E_{1-\varphi}((u/\lambda)^\gamma)$ in (\ref{one}) and~(\ref{two}),
where $E_p(z)$ denotes the generalized exponential integral with $p >
1$ and $z
\geq0$. However, for this present paper, we consider only $\gamma>
1$, as it is
difficult to envisage in our context a situation where an initially decreasing
hazard rate over time would arise.

The maximum likelihood estimates, $\hat{\theta}$, for $\theta$ can
now be
obtained by substituting these various expressions for the terms in
(\ref{like1}) into $L(\theta)=\prod_{i=1}^{n}L_i(\theta)$ and then
maximizing
$l(\theta)=\log L(\theta) = \sum_{i=1}^n l_i(\theta)$ over $\theta
$. Estimates
of the standard errors for $\hat{\theta}$ are obtained from inverting
the observed
information matrix, $-\partial^2l(\theta)/\partial\theta\,\partial
\theta^T$,
evaluated at $\hat{\theta}$.

%s3.2 #&#
\subsection{Hybrid pseudo score/profile likelihood approach}\label{sec3.2}
As an alternative to the full likelihood approach, a pseudo score method
based on inverse probability weights can be developed [\citet
{CookLawless2007}].
We assume that the incident population has $N$ individuals with
initiating event times\vspace*{1pt}
occurring in the period $d_1$ to $d_2$. The weighted pseudo score,
$U_1(\psi,\pi)$,
with $\psi^T=(\gamma,\beta^T)$, is constructed by weighting the
Weibull score contributions, $\partial l^{W}_{i}/\partial\psi$ for
selected subjects by $w_i = 1/p_i$ ($i=1,\ldots, n)$, where $p_i$ is the selection probability for subject $i$.
This weighted pseudo Weibull score, which has expectation zero, takes
the form
\begin{eqnarray*}
U_1(\psi, \pi) &=& \sum_{i=1}^N
{S_i} {w_i} \frac{\partial
l^{W}_{i}}{\partial\psi}
\\
&=& \sum_{i=1}^N\frac{S_i}{p_i}
\frac{\partial}{\partial\psi} \bigl[ \delta_i\log f_T(x_i
\mid z_i) + (1-\delta_i)\log\bigl(1-F_T(x_i
\mid z_i)\bigr) \bigr].
\end{eqnarray*}

For a selected study subject $i$ (i.e., $S_i=1$), $p_i$ is either
$\operatorname{pr}(0<
R_i < u_i=d_{2}-y_{i}|T_i=x_i)$ if $\delta_i=1$ or
$\operatorname{pr}(0< R_i < u_i|T_i \geq x_i)$ if $\delta_i=0$, with $x_i \leq
u_i$. The former probability expression
evaluates to $1$, as a subject who is observed to have experienced the
event of interest would have $t_i=x_i \leq u_i$ and since $T_i > R_i$ (by
Assumption~\ref{assumption1}), then, with probability $1$, $R_i < u_i$.
The latter probability expression is shown in the supplementary material [\citet{supp}] to be
$\{\operatorname{pr}(0 < R_i <u_i) - F_T(x_i \mid z_i)\}/\{1-F_T(x_i \mid
z_i)\}$, which is a function of $\theta$. These expressions are
derived under
the supposition that no further follow-up information on referred
individuals beyond
$d_2$, the close of the study, is available. This reflects the
situation in our
application. However, these expressions can be easily modified to take
account of
further follow-up information beyond the close of study, as shown in
the supplementary material [\citet{supp}]
for selected individuals with $\delta_i=0$ and $x_i > u_i$. The former
probability expression for
an uncensored selected individual $i$ is trivially $F_{R_i \mid
T_i}(u_i \mid x_i)$,
where $x_i$ can now be greater than $u_i$.\vspace*{1pt}

Estimation of $\theta^T=(\gamma,\beta^T,\pi^T)$ under this second approach
proceeds in two stages. First, $\psi$, the vector of Weibull shape and
regression parameters, is estimated by setting the pseudo score,
$U_1(\psi,\pi)$, to zero
and solving for $\psi$ with given $\{p_i\}$ to get the maximum
weighted pseudo score estimates of $\psi$.
Next, the inclusion probabilities $\{p_i\}$ for selected subjects with
$\delta_i=0$ are reevaluated at these maximum weighted pseudo score
estimates and
at the maximum profile likelihood estimate of $\pi$
obtained after maximizing $l(\theta)$ over $\pi$ with $\psi$ set in
(\ref{like1}) to its maximum weighted pseudo score estimates. These
two steps are
iterated until convergence of the estimates for $\theta$ to $\tilde
{\theta}$.
Initially the inclusion probabilities $\{p_i\}$ are all assumed to take the
value $1$ and, therefore, the initial estimate of $\psi$ is from the standard
(unweighted) Weibull regression model. This iterative estimation
procedure is similar to that used by
\citet{HardinHilbe2003} for longitudinal data, although, to
minimize efficiency loss,
we do not adopt their assumption of orthogonality of the estimating equations.

Estimated standard errors based on this approach can be obtained either through
a standard bootstrap procedure or determined based on Taylor series expansion
arguments applied to the set of unbiased estimating equations
$U_1(\psi,\pi)=0$ and $U_2(\psi,\pi) \equiv\partial l(\theta
)/\partial\pi= 0$.
Under appropriate regularity conditions, the asymptotic joint
distribution of
$((\tilde{\psi}-\psi)^T,(\tilde{\pi}-\pi)^T)$ is Gaussian with
expectation zero
and
variance--covariance matrix consistently estimated by the robust
sandwich matrix
$\Sigma\Lambda\Sigma^T$ evaluated at $\tilde{\theta}$, where
$\Sigma^{-1}$ is
\[
- \pmatrix{ \displaystyle\frac{\partial U_1}{\partial\psi^T} & \displaystyle\frac{\partial
U_1}{\partial\pi^T}
\vspace*{5pt}\cr
\displaystyle\frac{\partial U_2}{\partial\psi^T} &
\displaystyle\frac{\partial U_2}{\partial
\pi^T}}
\]
and\vspace*{1pt} $\Lambda= \sum_{\{i\dvtx S_i=1\}} U_{0i}U_{0i}^T$, where $U_{0i}^T=
(U_{1i}^T,U_{2i}^T)=(w_i(\theta)\,\partial l^{W}_{i}/\partial\psi^T,
\partial
l_i/\partial\pi^T)$ for $S_i=1$, with the dependency of $w_i(\theta
)$ on $\theta$
explicitly shown. With this extra notation, it is easily seen that
$U_1(\psi,\pi)=\sum_{i=1}^nU_{1i}$ and $U_2(\psi,\pi)=\sum
_{i=1}^nU_{2i}$, and
\begin{eqnarray*}
\frac{\partial U_1}{\partial\psi^T} &=& \sum_{i=1}^{n}
\biggl( \frac{\partial l^{W}_{i}}{\partial\psi} \frac{\partial w_i}{\partial\psi^T} + w_i
\frac{\partial^2 l^{W}_{i}}{\partial\psi\,\partial\psi^T} \biggr),\qquad
\frac{\partial U_1}{\partial\pi^T} = \sum
_{i=1}^{n} \frac
{\partial
l^{W}_{i}}{\partial\psi} \frac{\partial w_i}{\partial\pi^T},
\\
\frac{\partial U_2}{\partial\psi^T} &=& \sum_{i=1}^{n}
\frac
{\partial^2
l_i}{\partial\pi\,\partial\psi^T}\quad\mbox{and}\quad \frac{\partial
U_2}{\partial\pi^T} = \sum
_{i=1}^{n} \frac
{\partial^2
l_i}{\partial\pi\,\partial\pi^T}.
\end{eqnarray*}

%s3.3 #&#
\subsection{Simulation study: Consistency, efficiency and robustness considerations}\label{sec3.3}

To illustrate the performance of the proposed methods, in particular,
with regard to efficiency, bias and robustness, we conducted a small-scale
simulation with a design similar to that in \citet{FuTomBird2007}, \citet{FuTomBird2009}. We
performed 500 simulation runs and generated, in each of the runs, a
community sample size of $N=5000$. We considered three different
time-to-event distributions from which to
simulate our data. These were (i)~the Weibull, (ii) the gamma and (iii)~the log-normal. The parameter configurations for these
three distributions were
(i)~$\psi_W^T = (\gamma_W,\beta_0,\beta_1,\beta_2) = (4,
4\ta 6, -0\ta 03, -0\ta 4)$,
(ii) $\psi_G^T = (\gamma_G,\beta_0,\beta_1,\beta_2) = (
12\ta 71, 1\ta 96, -0\ta 03, -0\ta 4)$, and
(iii) $\psi_{LN}^T = (\sigma_{LN},\beta_0,\beta_1,\beta_2) =
(0\ta 275, 4\ta 464, -0\ta 03, -0\ta 4)$,
corresponding to the shape parameters, $\gamma_W$ and $\gamma_G$,
scale parameter, $\sigma_{LN}$, and the regression parameters
$\beta=(\beta_0,\beta_1,\beta_2)^T$ associated with the covariate
vector $z$ comprising of an intercept, a continuous variable, $z_1$,
generated from a log-normal distribution with location and scale
parameters taking the values 3 and 0\ta 3, respectively, and a
binary variable, $z_2$, generated from a Bernoulli distribution with
success probability of $1/3$. These covariates are included
through a log-linear regression model on the Weibull's and gamma's
scale parameters and a linear model on the log-normal's location parameter.
The parameters $\beta_0$, $\gamma_W$, $\gamma_G$ and $\sigma_{LN}$
were chosen to make the log baseline means from the regression models
corresponding
to the three distribution all equal to $4.50$. The truncation times (in
years), which are entry times for those selected, are generated from
the conditional
distribution proposed earlier with $\pi'=(\mbox{0\ta 1, 0\ta 06,
0\ta 12, 0\ta 24, 0\ta 48})^T$ and $\nu=(\mbox{0,
0\ta 5, 0\ta 625, 0\ta 75, 0\ta 875, 1})^T$. For
simplicity in interpretation
of the various simulation results to be presented,
we assume everyone in the community experienced the initiating event at the
same calendar date and those whose truncation time was less than
$d_0=15$ years
entered the referral cohort. Administrative right censoring of sampled
subjects occurred at
$c_0=15$ years from the calendar date of the initiating event. This was the
only type of censoring considered here. The parameters of the Weibull
distribution were informed by the data that arose from the Edinburgh
Royal Infirmary's hepatitis C virus liver clinic, which are analyzed
later.

\subsubsection*{Correct specification of the time-to-event distribution}
Table~\ref{tab1} presents the findings from the aforementioned
simulation. Both approaches
produce consistent estimates of the parameters from the Weibull model
and the
sampling mechanism when the time-to-event distribution was Weibull. As
expected, more efficient estimates of the shape
and regression parameters were obtained from the full likelihood
approach than
the hybrid pseudo score/profile likelihood approach. Similar estimated
standard errors
were obtained from both approaches for the corresponding estimates of
$\pi$. This perhaps reflects the near optimality of the hybrid
approach when
estimating $\pi$ since the relevant part of the full likelihood is
being used.

%
%t1 #&#
\begin{table}%[t]
\tabcolsep=0pt
\caption{Full and hybrid pseudo score/profiled Weibull likelihood
simulation results}\label{tab1}
\begin{tabular*}{\tablewidth}{@{\extracolsep{\fill}}@{}lcd{2.3}d{1.3}d{1.3}d{2.3}d{1.3}d{1.3}d{1.3}@{}}
\hline
& & \multicolumn{3}{c}{} & \multicolumn{3}{c}{\textbf{Hybrid pseudo score/}}\\
& & \multicolumn{3}{c}{\textbf{Full likelihood}} & \multicolumn{3}{c}{\textbf{profile likelihood}}\\[-6pt]
\multirow{2}{46pt}{\textbf{True distribution}}  & \multirow{2}{27pt}{\textbf{Para\-meters}}& \multicolumn{3}{c}{\hrulefill} & \multicolumn{3}{c}{\hrulefill}
\\
 &  & \multicolumn{1}{c}{\textbf{Mean}} &
\multicolumn{1}{c}{$\overline{\mbox{\textbf{SE}}}$} & \multicolumn{1}{c}{\textbf{ESE}} & \multicolumn{1}{c}{\textbf{Mean}} &
\multicolumn{1}{c}{$\overline{\mbox{\textbf{RSE}}}$} & \multicolumn{1}{c}{\textbf{ESE}} & \multicolumn{1}{c@{}}{\textbf{RE}}\\
\hline
Weibull & $\beta_0$ & 4\ta 635 & 0\ta 289 & 0\ta 287 & 4\ta 631 & 0\ta 386& 0\ta 397 & 0\ta 525 \\
& $\beta_1$ & -0\ta 031 & 0\ta 005 & 0\ta 006 & -0\ta 030 &0\ta 007 & 0\ta 007 & 0\ta 604 \\
& $\beta_2$ & -0\ta 405 & 0\ta 093 & 0\ta 090 & -0\ta 404 &0\ta 109 & 0\ta 112 & 0\ta 645 \\
& $\gamma_W$ & 4\ta 041 & 0\ta 345 & 0\ta 342 & 4\ta 062 & 0\ta 361& 0\ta 401 & 0\ta 729 \\
& $\pi_1$ & 0\ta 060 & 0\ta 020 & 0\ta 020 & 0\ta 061& 0\ta 018& 0\ta 019 & 1\ta 011 \\
& $\pi_2$ & 0\ta 120 & 0\ta 030 & 0\ta 031 & 0\ta 120& 0\ta 029& 0\ta 031 & 0\ta 998 \\
& $\pi_3$ & 0\ta 241 & 0\ta 041 & 0\ta 040 & 0\ta 240& 0\ta 040& 0\ta 040 & 1\ta 011 \\
& $\pi_4$ & 0\ta 480 & 0\ta 047 & 0\ta 047 & 0\ta 479& 0\ta 050& 0\ta 049 & 0\ta 929
\\[3pt]
Gamma & $\beta_0$ & 4\ta 378 & 0\ta 276 & 0\ta 307 & 4\ta 325 & 0\ta 277& 0\ta 335 & 0\ta 838 \\
& $\beta_1$ & -0\ta 028 & 0\ta 005 & 0\ta 006 & -0\ta 027 &0\ta 005 & 0\ta 006 & 0\ta 839 \\
& $\beta_2$ & -0\ta 405 & 0\ta 091 & 0\ta 091 & -0\ta 402 &0\ta 103 & 0\ta 111 & 0\ta 677 \\
& $\gamma_W$ & 6\ta 658 & 0\ta 772 & 0\ta 779 & 6\ta 821 & 0\ta 646& 0\ta 883 & 0\ta 779 \\
& $\pi_1$ & 0\ta 059 & 0\ta 022 & 0\ta 022 & 0\ta 060& 0\ta 019& 0\ta 022 & 1\ta 066 \\
& $\pi_2$ & 0\ta 114 & 0\ta 039 & 0\ta 041 & 0\ta 113& 0\ta 037& 0\ta 039 & 1\ta 080 \\
& $\pi_3$ & 0\ta 233 & 0\ta 054 & 0\ta 056 & 0\ta 231& 0\ta 053& 0\ta 054 & 1\ta 063 \\
& $\pi_4$ & 0\ta 482 & 0\ta 066 & 0\ta 071 & 0\ta 475& 0\ta 066& 0\ta 070 & 1\ta 046
\\[3pt]
Log-normal & $\beta_0$ & 4\ta 344 & 0\ta 269 & 0\ta 308 & 4\ta 338 & 0\ta 276& 0\ta 392 & 0\ta 617 \\
& $\beta_1$ & -0\ta 028 & 0\ta 004 & 0\ta 005 & -0\ta 028 &0\ta 005 & 0\ta 007 & 0\ta 671 \\
& $\beta_2$ & -0\ta 402 & 0\ta 094 & 0\ta 101 & -0\ta 416 &0\ta 113 & 0\ta 136 & 0\ta 554 \\
& $\gamma_W$ & 7\ta 919 & 0\ta 993 & 1\ta 071 & 8\ta 027 & 0\ta 767& 1\ta 230 & 0\ta 757 \\
& $\pi_1$ & 0\ta 062 & 0\ta 023 & 0\ta 025 & 0\ta 062& 0\ta 020& 0\ta 025 & 1\ta 007 \\
& $\pi_2$ & 0\ta 108 & 0\ta 042 & 0\ta 043 & 0\ta 109& 0\ta 040& 0\ta 041 & 1\ta 088 \\
& $\pi_3$ & 0\ta 225 & 0\ta 061 & 0\ta 062 & 0\ta 224& 0\ta 059& 0\ta 060 & 1\ta 053 \\
& $\pi_4$ & 0\ta 492 & 0\ta 076 & 0\ta 078 & 0\ta 484& 0\ta 076& 0\ta 075 & 1\ta 073\\
\hline
\end{tabular*}
\tabnotetext[]{}{Mean, average of the estimate; $\overline{\mbox{SE}}$, average of the
estimated standard error; ESE, empirical standard error; $\overline{\mbox
{RSE}}$, average of the estimated robust standard error;
RE, the empirical variance of the maximum likelihood estimator divided
by the empirical variance of the maximum hybrid pseudo score/profile
likelihood estimator.}
\end{table}

\subsubsection*{Misspecification of the time-to-event distribution}
Table~\ref{tab1} also shows the results when the true time-to-event
distributions are gamma and log-normal but the full likelihood and
hybrid pseudo score/profile likelihood approaches were fitted assuming
the time-to-event distribution was Weibull. Here we see that the impact
of this incorrect assumption for the time-to-event distribution is
negligible for the estimation of the regression parameters $\beta_1$
and $\beta_2$
and minor for the estimation of $\pi$. The estimates of the log
baseline mean under misspecification of the true gamma and log-normal
distributions by the Weibull [i.e.,
$\log(\Gamma(1+1/\gamma_W))+\beta_0$] were approximately $4.31$ and
$4.28$, respectively. As earlier mentioned, the true log baseline mean
is $4.50$. Therefore, for these
particular cases of misspecification there is underestimation of the
log baseline mean. This underestimation of the log baseline mean would
result in an underestimation
of the tail probabilities of the marginal population time-to-event
distribution, which would lead to underestimation of the population size.

%
%t2 #&#
\begin{table}
\tabcolsep=10pt
\caption{Full and hybrid pseudo score/profiled likelihood Weibull
simulation results under
a less parsimonious representation of the truncation time conditional
distribution}\label{tab2}
\begin{tabular*}{\tablewidth}{@{\extracolsep{\fill}}ld{2.3}d{1.3}d{1.3}d{2.3}d{1.3}d{1.3}@{}}
\hline
& \multicolumn{3}{c}{} & \multicolumn{3}{c@{}}{\textbf{Hybrid pseudo score/}}\\
& \multicolumn{3}{c}{\textbf{Full likelihood}} & \multicolumn{3}{c@{}}{\textbf{profile likelihood}}\\[-6pt]
& \multicolumn{3}{c}{\hrulefill} & \multicolumn{3}{c@{}}{\hrulefill}\\
\textbf{Parameters} & \multicolumn{1}{c}{\textbf{Mean}} & \multicolumn{1}{c}{$\overline{\mbox{\textbf{SE}}}$}
& \multicolumn{1}{c}{\textbf{ESE}} & \multicolumn{1}{c}{\textbf{Mean}} & \multicolumn{1}{c@{}}{$\overline{\mbox{\textbf{RSE}}}$}
& \multicolumn{1}{c@{}}{\textbf{ESE}} \\
\hline
$\beta_0$ & 4\ta 606 & 0\ta 255 & 0\ta 258 & 4\ta 604& 0\ta 238& 0\ta 252 \\
$\beta_1$ & -0\ta 030 & 0\ta 005 & 0\ta 005 & -0\ta 030 &0\ta 005 & 0\ta 005 \\
$\beta_2$ & -0\ta 400 & 0\ta 085 & 0\ta 087 & -0\ta 400 &0\ta 084 & 0\ta 086 \\
$\gamma$ & 4\ta 057 & 0\ta 394 & 0\ta 385 & 4\ta 056 &0\ta 363& 0\ta 380 \\
$\pi_1$ & 0\ta 125 & 0\ta 027 & 0\ta 026 & 0\ta 125 &0\ta 026& 0\ta 026 \\
$\pi_2$ & 0\ta 125 & 0\ta 030 & 0\ta 030 & 0\ta 125 &0\ta 030& 0\ta 031 \\
$\pi_3$ & 0\ta 127 & 0\ta 032 & 0\ta 031 & 0\ta 127 &0\ta 031& 0\ta 031 \\
$\pi_4$ & 0\ta 125 & 0\ta 032 & 0\ta 032 & 0\ta 125 &0\ta 032& 0\ta 031 \\
\hline
\end{tabular*}
\tabnotetext[]{}{Mean, average of the estimate; $\overline{\mbox{SE}}$, average of the
estimated standard error; ESE, empirical standard error; $\overline{\mbox{RSE}}$, average
of the estimated robust standard error.}
\end{table}

\subsubsection*{Misspecification of the referral mechanism}
To investigate the relative robustness of the proposed mixture of
uniforms for the conditional distribution of the truncation
times given the time to event, we began by rerunning our simulation
study as before, except with the conditional
distribution of the truncation time now generated from a single uniform
distribution in the interval zero to the true time to event
instead of the five-component mixture of uniforms. However, the \textit{less parsimonious}
five-component mixture of uniforms,
with $\nu=(0, 0\ta 5, 0\ta 625, 0\ta 75, 0\ta
875, 1)^T$, was assumed as the working conditional distribution
of the truncation times when fitting the full likelihood and hybrid
approaches to the simulated data at each simulation run.
The results are shown in Table~\ref{tab2}. The less parsimonious
working conditional distribution for the truncation times has no
apparent impact, as
would be expected, on consistent estimation of the regression and shape
parameters of the Weibull distribution.
This suggests that finer partitions of $\nu$ than needed do not impact
on consistency of estimated regression and shape parameters, although
may inflate the standard errors of these estimates and the mixture
probability estimates.

%
%t3 #&#
\begin{table}%[t]
\tabcolsep=5pt
\caption{Full and hybrid pseudo score/profiled Weibull likelihood
simulation results under
misspecification of the truncation time conditional distribution by a
cruder partitioning}\label{tab3}
\begin{tabular*}{\tablewidth}{@{\extracolsep{\fill}}ld{2.3}d{1.3}d{1.3}d{2.3}d{1.3}d{1.3}@{}}
\hline
& \multicolumn{3}{c}{} & \multicolumn{3}{c@{}}{\textbf{Hybrid pseudo score/}}\\
& \multicolumn{3}{c}{\textbf{Full likelihood}} & \multicolumn{3}{c@{}}{\textbf{profile likelihood}}\\[-6pt]
& \multicolumn{3}{c}{\hrulefill} & \multicolumn{3}{c@{}}{\hrulefill}\\
\textbf{Parameters} & \multicolumn{1}{c}{\textbf{Mean}} & \multicolumn{1}{c}{$\overline{\mbox{\textbf{SE}}}$}
& \multicolumn{1}{c}{\textbf{ESE}} & \multicolumn{1}{c}{\textbf{Mean}} & \multicolumn{1}{c@{}}{$\overline{\mbox{\textbf{RSE}}}$}
& \multicolumn{1}{c@{}}{\textbf{ESE}} \\
\hline
$c_0=d_0=15$ & & & & & & \\
$\beta_0$ & 3\ta 841 & 0\ta 133 & 0\ta 156 & 4\ta 790& 0\ta 383 & 0\ta 388 \\
$\beta_1$ & -0\ta 018 & 0\ta 003 & 0\ta 004 & -0\ta 033 & 0\ta 007 & 0\ta 007 \\
$\beta_2$ & -0\ta 211 & 0\ta 052 & 0\ta 056 & -0\ta 427 & 0\ta 107 & 0\ta 108 \\
$\gamma$ & 5\ta 231 & 0\ta 391 & 0\ta 470 & 4\ta 048 &0\ta 341 & 0\ta 392 \\
$\pi_1$ & 0\ta 244 & 0\ta 030 & 0\ta 028 & 0\ta 252 &0\ta 032 & 0\ta 037 \\
$\pi_2$ & 0\ta 132 & 0\ta 030 & 0\ta 029 & 0\ta 156 &0\ta 035 & 0\ta 035 \\
$\pi_3$ & 0\ta 169 & 0\ta 032 & 0\ta 033 & 0\ta 193 &0\ta 037 & 0\ta 038 \\
$\pi_4$ & 0\ta 214 & 0\ta 036 & 0\ta 038 & 0\ta 244 &0\ta 041 & 0\ta 043
\\[3pt]
$c_0=d_0=20$ & & & & & & \\
$\beta_0$ & 4\ta 224 & 0\ta 100 & 0\ta 112 & 4\ta 654& 0\ta 158 & 0\ta 174 \\
$\beta_1$ & -0\ta 023 & 0\ta 002 & 0\ta 003 & -0\ta 031 & 0\ta 003 & 0\ta 004 \\
$\beta_2$ & -0\ta 290 & 0\ta 039 & 0\ta 041 & -0\ta 417 & 0\ta 059 & 0\ta 062 \\
$\gamma$ & 4\ta 565 & 0\ta 227 & 0\ta 253 & 4\ta 020 &0\ta 212 & 0\ta 230 \\
$\pi_1$ & 0\ta 199 & 0\ta 018 & 0\ta 018 & 0\ta 196 &0\ta 019 & 0\ta 019 \\
$\pi_2$ & 0\ta 146 & 0\ta 019 & 0\ta 019 & 0\ta 153 &0\ta 020 & 0\ta 020 \\
$\pi_3$ & 0\ta 184 & 0\ta 020 & 0\ta 020 & 0\ta 194 &0\ta 022 & 0\ta 022 \\
$\pi_4$ & 0\ta 230 & 0\ta 023 & 0\ta 023 & 0\ta 247 &0\ta 024 & 0\ta 025
\\[3pt]
$c_0=d_0=30$ & & & & & & \\
$\beta_0$ & 4\ta 527 & 0\ta 051 & 0\ta 048 & 4\ta 603& 0\ta 056 & 0\ta 054 \\
$\beta_1$ & -0\ta 028 & 0\ta 001 & 0\ta 001 & -0\ta 030 & 0\ta 002 & 0\ta 001 \\
$\beta_2$ & -0\ta 374 & 0\ta 023 & 0\ta 023 & -0\ta 402 & 0\ta 025 & 0\ta 025 \\
$\gamma$ & 4\ta 114 & 0\ta 112 & 0\ta 111 & 4\ta 013 &0\ta 109 & 0\ta 112 \\
$\pi_1$ & 0\ta 151 & 0\ta 009 & 0\ta 010 & 0\ta 150 &0\ta 009 & 0\ta 010 \\
$\pi_2$ & 0\ta 152 & 0\ta 010 & 0\ta 010 & 0\ta 152 &0\ta 010 & 0\ta 011 \\
$\pi_3$ & 0\ta 197 & 0\ta 011 & 0\ta 011 & 0\ta 199 &0\ta 011 & 0\ta 011 \\
$\pi_4$ & 0\ta 245 & 0\ta 012 & 0\ta 013 & 0\ta 249 &0\ta 013 & 0\ta 013 \\
\hline
\end{tabular*}
\tabnotetext[]{}{Mean, average of the estimate; $\overline{\mbox{SE}}$, average of the
estimated standard error; ESE, empirical standard error; $\overline{\mbox
{RSE}}$, average of the estimated robust standard error.}
\end{table}

To explore the impact of misspecification due to the incorrect
partitioning of $\nu$,
we again repeated the simulation study but now allowing the truncation
times to be generated from an eight-component
mixture of uniform conditional distribution, with $\nu=(0,
0\ta 125,\ldots, 0\ta 875, 1)^T$ and
$\pi' = (0 \ta 025, 0\ta 05, 0\ta 1, 0\ta 1,
0\ta 125, 0\ta 15, 0\ta 2, 0\ta 25)^T$, mimicking a
strong preference
for referrals to occur in the last half of individuals' incubation
period. Additionally, we considered three scenarios for recruitment and
administrative censoring,
which reflected an increasing number of individuals referred and
observed experiencing the event of interest and thus providing
more information to the analysis: (i) $c_0=d_0=15$; (ii) $c_0=d_0=20$;
and (iii) $c_0=d_0=30$. We fitted the simulated
data sets assuming the working five-component conditional truncation
time distribution mentioned earlier, which is based
on a \textit{coarser} partitioning of $\nu$ than the true generating
mechanism. The results are shown in Table~\ref{tab3}. Here, we
see a noticeable negative impact of misspecification, from a cruder
partitioning of $\nu$, on the estimates
of the regression parameters and shape parameter (and mixture
probabilities) when using the full likelihood approach (i.e., bias), which
diminishes as the amount of information from the sample increases (as
reflected by the diminishing standard errors). There is, however,
no noticeable bias observed in the estimates of the regression
parameters and shape parameter obtained using the hybrid approach.
Moreover, there
is no apparent bias, under this hybrid approach, in the mixture
probabilities, except for $\pi_1$, where the bias decreases as the
information content
increases. This result suggests that the hybrid approach is
significantly more robust to misspecification than the full likelihood
approach under various
data scenarios, but at the cost of being less efficient in general.
This perhaps is due to the hybrid approach being a two-stage method.

%s3.4 #&#
\subsection{Application to Edinburgh Royal Infirmary's hepatitis C virus liver clinic}\label{sec3.4}
The hepatitis C virus epidemic is a major public health concern in the
UK and across the
world. To project national hepatitis C virus burden, unbiased
estimation of the
progression rate from infection to liver cirrhosis is required for the whole
community of hepatitis C viral infected individuals. Often, however,
the available
data on progression to cirrhosis are from a biased sample
of the population of interest. In the application we consider here, the
data on
$387$ individuals infected with the hepatitis C virus prior to 2000 (i.e.,
within the calendar period 1950 to 2000) arose from the Edinburgh Royal
Infirmary's hepatitis C virus liver clinic, a tertiary referral
hospital clinic
whereby patients with more rapid disease progression, or symptomatic
disease, would
be preferentially referred, with referral increasingly likely to be
closer to onset
of cirrhosis. Thus, it is important to account for this
outcome-dependent recruitment
when analyzing these data so as to provide realistic estimates of the
progression
rates and the effects of risk factors on time to cirrhosis from
infection for the
Edinburgh's community (of unknown size) of hepatitis C virus-infected
individuals.

To investigate the pattern of referral over patients' cirrhosis
incubation period, we
model the referral time $R$ given the cirrhosis time $T$ as coming from the
probability density function
\begin{eqnarray*}
f_{R \mid T}(r \mid t) &=& \sum_{j=0}^1
\frac{4\pi_j}{t}I\biggl(\frac{j}{4} < \frac{r}{t} \leq
\frac{j+1}{4}\biggr) + \sum_{j=2}^5
\frac{8\pi_j}{t}I\biggl(\frac{j+2}{8} < \frac{r}{t} \leq
\frac{j+3}{8}\biggr),
\end{eqnarray*}
where $\{\pi_j\dvtx  j=0,\ldots,5 \}$ are
the unknown mixture probabilities, summing to $1$, that are required
to be estimated. This distribution is chosen because of the clinical
belief that
patients are more likely to be referred in the last half of their cirrhosis
incubation period and we therefore decided to model this half in more detail.

Furthermore, we assume that, for the $i$th hepatitis C virus patient in the
community, the time to cirrhosis from known infection time follows a Weibull
distribution with unknown shape and scale parameters, $\gamma$ and
$\lambda_i$, respectively. The scale parameter, $\lambda_i$, is related to the $i$th
patient's continuous and binary explanatory variables, age at hepatitis
C viral
infection ($x_{1i}$) and excessive alcohol consumption ($x_{2i}$), through
the relationship $\log\lambda_i=\beta_0+\beta_1x_{1i}+\beta
_2x_{2i}$, where
$\beta^T=(\beta_0,\beta_1,\beta_2)$ is the vector of regression parameters.

%
%t4 #&#
\begin{table}
\tabcolsep=0pt
\caption{Full and hybrid pseudo score/profile likelihood results for
time to cirrhosis from hepatitis C virus (HCV) infection data from
Edinburgh Royal Infirmary's liver clinic prior to 2000}\label{tab4}
\begin{tabular*}{\tablewidth}{@{\extracolsep{\fill}}ld{2.3}d{1.3}d{2.3}d{1.3}d{1.3}@{}}
\hline
& \multicolumn{2}{c}{} & \multicolumn{3}{c@{}}{\textbf{Hybrid pseudo score/}}\\
& \multicolumn{2}{c}{\textbf{Full likelihood}} & \multicolumn{3}{c@{}}{\textbf{profile likelihood}}\\[-6pt]
& \multicolumn{2}{c}{\hrulefill} & \multicolumn{3}{c@{}}{\hrulefill}
\\
\textbf{Parameters} & \multicolumn{1}{c}{\textbf{Estimate}} & \multicolumn{1}{c}{\textbf{s.e.}}
& \multicolumn{1}{c}{\textbf{Estimate}} & \multicolumn{1}{c}{\textbf{Robust s.e.}} & \multicolumn{1}{c@{}}{\textbf{Bootstrap s.e.}}
\\
\hline
$\beta_0$ & 4\ta 410 & 0\ta 123 & 4\ta 380 & 0\ta 174& 0\ta 181\\
$\beta_1$ & -0\ta 022 & 0\ta 004 & -0\ta 023 & 0\ta 004 & 0\ta 004\\
$\beta_2$ & -0\ta 521 & 0\ta 082 & -0\ta 494 & 0\ta 109 & 0\ta 111\\
$\gamma$ & 4\ta 948 & 0\ta 408 & 5\ta 256 & 0\ta 441 &0\ta 467\\
$\pi_1$ & 0\ta 096 & 0\ta 020 & 0\ta 093 & 0\ta 014 &0\ta 015\\
$\pi_2$ & 0\ta 065 & 0\ta 055 & 0\ta 060 & 0\ta 050 &0\ta 046\\
$\pi_3$ & 0\ta 126 & 0\ta 076 & 0\ta 139 & 0\ta 070 &0\ta 065\\
$\pi_4$ & 0\ta 064 & 0\ta 036 & 0\ta 062 & 0\ta 035 &0\ta 036\\
$\pi_5$ & 0\ta 639 & 0\ta 055 & 0\ta 635 & 0\ta 051 &0\ta 041
\\[3pt]
Log-likelihood & \multicolumn{2}{c}{$-$1259.78} \\
N (bootstrap IQR) & & & \multicolumn{1}{c}{4196} & \multicolumn{1}{c}{(3414--5139)} \\
\hline
\end{tabular*}
\tabnotetext[]{}{s.e., standard error; N, estimated size of Edinburgh's hepatitis C virus
community prior to 2000; IQR, inter-quartile range; $\beta_1$ and
$\beta_2$, regression coefficients corresponding to age at HCV infection and excessive
alcohol consumption status.}
\end{table}

Table~\ref{tab4} shows the results obtained on fitting the Edinburgh
Royal Infirmary data using both the full likelihood and hybrid pseudo
score/profile
likelihood approaches. The bootstrap standard errors are obtained from
a bootstrap sample of $500$. Relatively similar regression parameter estimates
and corresponding estimated standard errors are obtained from the two
approaches. The belief by clinicians that referral was more likely in
the last half
of the cirrhosis period is borne out with about $90\%$ of infected individuals
estimated to be referred then. Strikingly, about $64\%$ of infected
individuals are estimated
to have been referred in the last one eighth of their cirrhosis period.
On repeating the analysis
with a cruder representation of the referral mechanism based on
partitioning $\nu$ only into halves produced
fairly similar regression estimates under both approaches (data not
shown) to those in Table~\ref{tab4}. However, for the
more variable Weibull shape parameter, there are noticeable differences
in the estimates from this cruder referral mechanism
to those previously obtained, in particular, for the full likelihood
approach. The estimates (with standard errors) of the shape parameter
are now
3\ta 833 (0\ta 360) and 4\ta 755 (0\ta 400) for the
full likelihood and hybrid approaches, respectively. Additionally, the
estimates of the probability of being referred in the last half of the
incubation period, obtained assuming the cruder referral mechanism,
are roughly equal under the two approaches but now calculated to be
approximately $98\%$ as opposed to the $90\%$ previously
estimated.\looseness=1

From the hybrid method, we obtain an estimate (bootstrap inter-quartile
range) of 4196 (3414--5139) infected individuals in
Edinburgh's hepatitis C virus community prior to 2000, through the
summation of the inverse probability weights. Both older
age at onset of infection and excessive alcohol consumption are found
statistically significantly to increase the rate of progression
to cirrhosis. The corresponding relative risk estimates (with 95\%
confidence intervals) for age at infection onset and for excessive
alcohol consumption are
1\ta 13 (1\ta 09, 1\ta 18) and 13\ta 4 (5\ta 1,
35\ta 3), respectively.
The (inverse probability weighted) estimates of the mean age at HCV
infection (with standard deviation) and the proportion
consuming excessive alcohol in the Edinburgh HCV community prior to
2000 are 20\ta 3 (6\ta 3) years and 6\ta 7\%, respectively.
For comparison,
the mean age at HCV infection (with standard deviation) and the
proportion consuming excessive alcohol from the Edinburgh Royal
Infirmary clinic data are
22\ta 4 (9\ta 8) years and 30\%, respectively. There is a
striking difference in the community's and clinic's estimates of
the proportion consuming excessive alcohol.

An estimated marginal 30-year progression rate (with sampling
uncertainty) to cirrhosis from infection in the Edinburgh HCV community
can also be calculated through a fast
parametric bootstrap-like approach [\citet{AalenFarewell1997}]. Here
we repeatedly sample $\theta^T=(\gamma,\beta^T,\pi^T)$ from the
asymptotic distribution of
$\tilde{\theta}$, specified by a multivariate normal distribution
with mean vector and variance--covariance matrix given by $\tilde
{\theta}$
and the robust sandwich matrix, $\Sigma\Lambda\Sigma^T$ evaluated at
${\tilde{\theta}}$.
For each of these sampled parameter vectors, we define a corresponding
hypothetical Edinburgh HCV community (prior to 2000) that can be
entirely followed up to
cirrhosis. The size, mean and standard deviation of the age at HCV
infection and the excessive alcohol consumption proportion for each of
these hypothetical
communities are calculated by applying the inverse probability weights,
calculated using the sampled $\theta$'s, to the Edinburgh Royal
Infirmary data. The
communities' cirrhosis data can be constructed by generating the times
to cirrhosis from the proposed Weibull model using the sampled
regression and
shape parameters, after first simulating the explanatory variables, age
at HCV infection and alcohol consumption status, for the created communities.
We assume that the age at HCV infection and alcohol consumption status
distributions are independent from one another and are log-normal and Bernoulli,
respectively, with mean and standard deviation (for the log normal) and
excessive alcohol consumption proportion parameters arising from the
application of the
sampled $\theta$, through the generated inverse probability weights,
to the Edinburgh Royal Infirmary data. Our assumption of marginal
independence is based on an
estimated Pearson's correlation between age at HCV infection and
excessive alcohol consumption status of 0\ta 012 in the actual
collected data, which we
do not anticipate to change dramatically when translated to the community.

\begin{figure}

\includegraphics{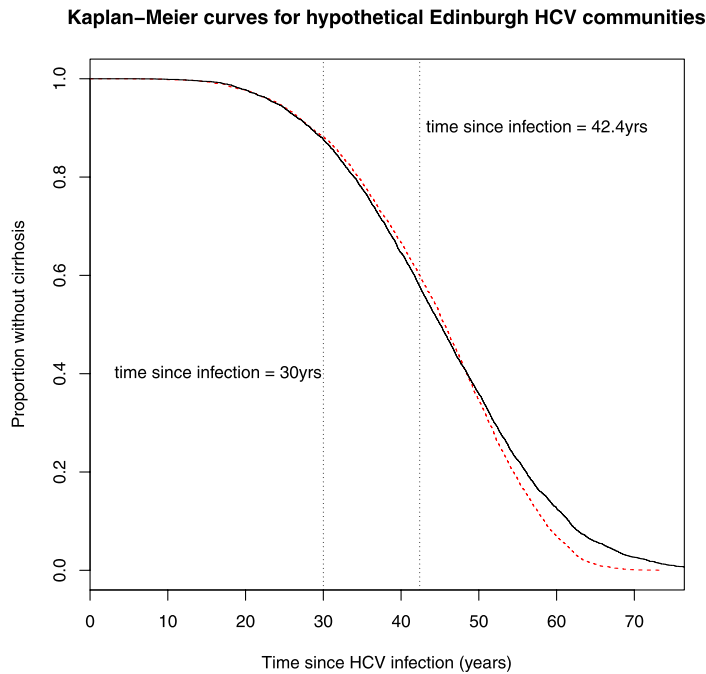}

\caption{Marginal Kaplan--Meier curves for hypothetical Edinburgh HCV
communities derived under the assumption
that the time-to-cirrhosis distribution is either Weibull (solid line)
or generalized gamma (dashed line).
The vertical dotted lines correspond to time from infection of $30$
years and the last observed time in the
Edinburgh liver clinic data of $42.4$ years, which corresponds to an
(uncensored) cirrhotic event.}\vspace*{-2pt}\label{fig2}
\end{figure}

The application of this fast parametric bootstrap-like approach, over
500 runs, gave a mean 30-year progression rate to cirrhosis in the
hypothetical communities
of 14\% with a standard deviation of 6\% and a 95\% range of 6\% to 29\%. Figure~\ref{fig2} provides an example of a marginal Kaplan--Meier
curve for one hypothetical Edinburgh
community generated at the maximum weighted pseudo score estimates.

%
%f2 #&#

The estimated 30-year Kaplan--Meier progression rate (with 95\%
confidence interval) to cirrhosis based on the actual Edinburgh Royal
Infirmary data, ignoring the outcome-dependent referral and
left truncation, is 42\% (31\%, 52\%). The corresponding conditional
Kaplan--Meier estimate (conditioning on not experiencing cirrhosis at
least roughly 1 year after infection), assuming that
the Edinburgh Royal Infirmary data is a length-biased sample of the
Edinburgh HCV community, is 86\% with a 95\% confidence interval of
(75\%, 92\%). Both of these standard estimates dramatically
overestimate the 30-year progression rate, as they do not account for
the correlation between the referral time and the time to cirrhosis of
a referred patient.

To check the robustness of our findings for the Edinburgh data, we
implemented our two approaches replacing the assumption of a Weibull
time-to-event distribution with
the generalized gamma distribution (see the supplementary material [\citet{supp}] for its formulation),
which has one extra parameter and includes the Weibull, gamma and
log-normal all as special cases. Although a likelihood ratio
test on 1 degree of freedom ($p=0.02$) rejected the Weibull in favor of
the generalized gamma, the maximum likelihood estimates for the
regression parameters of interest were similar to those
previously obtained ($\hat{\beta}_1=-0.021$ and $\hat{\beta
}_2=-0.546$) and the estimate of the proportion of infected individuals
referred to in the last one eighth of their cirrhosis period was
again $64\%$. Furthermore, the estimated mean 30-year progression rate
to cirrhosis was similar. On closer inspection, we found that the differences
between the assumption of a generalized gamma and that of a Weibull for
the time-to-event distribution was noticeable only in the upper tails
of the estimated marginal
time-to-event distributions of the Edinburgh HCV community, past the
actual largest observed time to cirrhosis (i.e., an uncensored event of
$42.4$ years)
seen from the Edinburgh Royal Infirmary's liver clinic. Similar to the
marginal Kaplan--Meier curve presented under the assumption that the
time-to-event distribution is Weibull, Figure~\ref{fig2} also displays
the equivalent Kaplan--Meier curve under the assumption of the
generalized gamma, and thus provides an illustration of the discrepancy
between curves being evident in the upper tail beyond $42.4$ years.

%s4 #&#
\section{Discussion}\label{sec4}\label{discuss}
A weighted pseudo score method is commonly suggested for handling
response-biased observations, where specifying the full likelihood
is difficult. Provided that the inverse probability weights can be
consistently estimated, then consistency of regression parameter
estimates will be achieved using this approach. However, if the full
likelihood is available, then it is generally preferable to use it
to estimate the parameters of interest, as these estimates will be more
efficient than those obtained from the weighted pseudo score method. This
preference for the full likelihood over the weighted pseudo score
method also holds when the time-to-event distribution is misspecified.
In the context of
misspecification, we would advocate fitting the more flexible
generalized gamma time-to-event distribution (or an alternative such as
a semi-parametric piecewise exponential-type
distribution), instead of the Weibull, in order to get better estimates
of the marginal progression rates. Nevertheless, it is worth noting
that a by-product of the weighted pseudo score approach,
which makes it appealing, is the straightforward estimation of the
total incident population size. This is not directly available
(although calculable) from the full likelihood approach.
In public health terms, estimation of the total number of HCV carriers
and the ``true'' impact of covariates on HCV progression are key.

In the informative entry time problem addressed here, we were able to
develop both a full likelihood approach and a hybrid two-stage pseudo
score/profile likelihood
approach for outcome-dependent referral where sampling is dependent on
the residual fraction of time remaining to develop the event. Under
correct specification of the referral mechanism, we found that the full
likelihood approach was indeed more efficient than the hybrid approach in
the estimation of the regression parameters of interest and the shape
parameter. The former approach, however, appeared to be more
susceptible to bias if the outcome-dependent
referral mechanism was misspecified through a coarser representation
and the ``information content'' of the data (in terms of number of referrals,
number of events and length of follow-up) was low. In the situation
where the ``information content'' is considered to be
relatively high, it perhaps would be more appealing to adopt the hybrid
method over the full likelihood approach, as it could be significantly more
robust and the decrease in the resulting efficiency may still be
acceptable. In general, we would recommend that when using either
approach, and, in particular,
the full likelihood one, analysts should begin by specifying a
reasonably fine partitioning of the $\nu$ which can then be refined to
obtain a
more parsimonious representation of the referral mechanism. This
strategy would allow checking for sensitivity/robustness to
misspecification of the referral mechanism.
However, convergence issues may arise if the partitioning is too fine
or if the selection probability for a subject is too small. We have not
investigated these
convergence issues here.

The application of these methods to data from the Edinburgh Royal
Infirmary's hepatitis C virus liver clinic allowed us to characterize
realistically the extent of Edinburgh's HCV epidemic prior to 2000 in
terms of progression rate to cirrhosis and the impact of alcohol
consumption and
age at HCV infection on this progression. Standard survival analysis
methods severely overestimated the 30-year progression rate and
underestimated the relative
risks for the explanatory variables.

In our present analysis of the Edinburgh HCV data, we assumed that the
time of infection was known. This simplifying assumption was thought
reasonable in our case, since even when the
times of HCV infection in the Edinburgh liver clinic were uncertain,
this uncertainty tended to be only in the determination of the exact
date of infection within a calendar year or two.
As the mean incubation period to cirrhosis is several orders of
magnitude greater than the size of this interval, we expect that, for
the analysis of our data set, the added uncertainty in estimation
due to this imprecision in timing of infection will be inconsequential.
However, in other applications where the timing of the initiation event
(e.g., cancer or HIV infection onset) is known only to
within an interval, which may be quite large, and where either the mean
time to the event of interest (e.g., death or AIDS) or the mean
follow-up time are not of an appreciably long enough length
compared to the mean width\vadjust{\goodbreak} of these intervals, the implications for
analyses of assuming-known initiation time (e.g., by choosing the
mid-point of the interval) can be major.
\citet{StruthersFarewell1989} discuss an approach to account
for unknown onset times, when the time of onset is known only to be in
an interval, say, $(a,b)$. This approach requires the specification
of a density, say, $g$, for the time of infection (e.g., a uniform
distribution) over the interval $(a,b)$. The likelihood to be optimized
over the parameters then takes the form
$\prod_{i=1}^{n}\int_{a_i}^{b_i}L_i(\theta)g_i(y;\tau)\,dy$, where
$L_i(\theta)$ is the likelihood contribution from the $i$th subject,
given known infection time, and the density, $g_i(\cdot)$, for this
subject's time of infection may be specified up to an unknown parameter
vector, $\tau$. Therefore, it can be seen that this approach can be
adapted to our situation where the sampling is dependent on the
residual fraction of time left to developing the event of interest and
the onset time is known only to within an interval. However, careful
thought is required on the most appropriate form for $g$.
For example, in HCV studies where the majority of subjects are
injecting drug users and when time of HCV infection is unknown, there
is evidence to suggest that infection
occurs earlier in a subject's injecting career [\citet
{Hutchinson2005}, \citet{Hagan2008}, \citet{Deangelis2009}].

Future application of these methods to the HCV epidemic in Scotland,
more generally, is planned with Health Protection Scotland. Health
Protection Scotland has
developed a clinical database on referrals of HCV patients to liver
clinics across all regions of Scotland. Application of our methodology
should provide regional estimates for
the number of HCV carriers in Scotland and will allow us to examine if
the ``true'' impact of covariates (such as age at HCV infection and
heavy alcohol use) are stable
across regions although the covariate distribution may differ between
regions.\vspace*{-1pt}

\section*{Acknowledgments}
We would like to acknowledge the anonymous referees and Editor for
their insightful comments and suggestions which have helped improve the
paper.\vspace*{-2pt}

% AOS,AOAS: If there are supplements please fill:
%
\begin{supplement}%[id=suppA]
\stitle{Appendix: Derivations of the expressions based on the generalized gamma and mixture of uniforms}
\slink[doi]{10.1214/14-AOAS725SUPP}%[doi,text={...}] - jei reikia %suskaldyti doi
\sdatatype{.pdf}
\sfilename{AOAS725\_supp.pdf}
\sdescription{Proofs of the various expressions required in the
constructing of the likelihood and pseudo score based on the assumption
that the time-to-event distribution is from a generalized gamma
distribution and the conditional referral distribution is a mixture of
independent uniforms.}
\end{supplement}

% zodis "Acknowledgments" paliekamas pagal autoriu

%suskaldyti doi

% imsref loaded by linak, 2014-03-14 14:56:22
%
% imsref loaded by linak, 2014-03-14 15:18:22

\printaddresses


\begin{thebibliography}{28}

%b1 #&#
\bibitem[\protect\citeauthoryear{Aalen et~al.}{1997}]{AalenFarewell1997}
%
\begin{barticle}[pbm]
\bauthor{\bsnm{Aalen},~\bfnm{O.~O.}\binits{O.~O.}},
\bauthor{\bsnm{Farewell},~\bfnm{V.~T.}\binits{V.~T.}},
\bauthor{\bsnm{Angelis},~\bfnm{D.~De}\binits{D.~D.}},
\bauthor{\bsnm{Day},~\bfnm{N.~E.}\binits{N.~E.}} \AND
\bauthor{\bsnm{Gill},~\bfnm{O.~N.}\binits{O.~N.}}
(\byear{1997}).
\btitle{A~Markov model for HIV disease progression including the effect
of HIV diagnosis and treatment: Application to AIDS prediction in
England and Wales}.
\bjournal{Stat. Med.}
\bvolume{16}
\bpages{2191--2210}.
\bid{issn={0277-6715},
pii={10.1002/(SICI)1097-0258(19971015)16:19<2191::AID-SIM645>3.0.CO;2-5},
pmid={9330428}}
\end{barticle}
%
\bptok{imsref}%
% NOT OUTPUTED:
% issn = 0277-6715
% number = 19
% fjournal = Statistics in medicine
\endbibitem

%b2 #&#
\bibitem[\protect\citeauthoryear{Andersen et~al.}{1993}]{AndersenBorganGill1993}
%
\begin{bbook}[mr]
\bauthor{\bsnm{Andersen},~\bfnm{Per~Kragh}\binits{P.~K.}},
\bauthor{\bsnm{Borgan},~\bfnm{{\O}rnulf}\binits{{\O}.}},
\bauthor{\bsnm{Gill},~\bfnm{Richard~D.}\binits{R.~D.}} \AND
\bauthor{\bsnm{Keiding},~\bfnm{Niels}\binits{N.}}
(\byear{1993}).
\btitle{Statistical Models Based on Counting Processes}.
\bpublisher{Springer},
\blocation{New York}.
\bid{doi={10.1007/978-1-4612-4348-9}, mr={1198884}}
\end{bbook}
%
\bptok{imsref}%
% NOT OUTPUTED:
% isbn = 0-387-97872-0
% url = http://dx.doi.org/10.1007/978-1-4612-4348-9
% fpage = xii+767
\endbibitem

%b3 #&#
\bibitem[\protect\citeauthoryear{Brookmeyer}{2005}]{Brookmeyer2005}
%
\begin{bincollection}[author]
\bauthor{\bsnm{Brookmeyer},~\bfnm{R.}\binits{R.}}
(\byear{2005}).
\btitle{Biased sampling of cohorts}.
In \bbooktitle{Encyclopedia of Biostatistics},
\bedition{2nd} ed.
(\beditor{\bfnm{P.}\binits{P.}~\bsnm{Armitage}} \AND
\beditor{\bfnm{T.}\binits{T.}~\bsnm{Colton}}, eds.)
\bpages{427--439}.
\bpublisher{Wiley},
\blocation{New York}.
\end{bincollection}
%
\bptok{imsref}%
\endbibitem

%b4 #&#
\bibitem[\protect\citeauthoryear{Cook and Lawless}{2007}]{CookLawless2007}
%
\begin{bbook}[author]
\bauthor{\bsnm{Cook},~\bfnm{R.~J.}\binits{R.~J.}} \AND
\bauthor{\bsnm{Lawless},~\bfnm{J.~F.}\binits{J.~F.}}
(\byear{2007}).
\btitle{The Statistical Analysis of Recurrent Events}.
\bpublisher{Springer},
\blocation{Berlin}.
\end{bbook}
%
\bptok{imsref}%
\endbibitem

%b5 #&#
\bibitem[\protect\citeauthoryear{Copas and Farewell}{2001}]{CopasFarewell2001}
%
\begin{barticle}[pbm]
\bauthor{\bsnm{Copas},~\bfnm{A.~J.}\binits{A.~J.}} \AND
\bauthor{\bsnm{Farewell},~\bfnm{V.~T.}\binits{V.~T.}}
(\byear{2001}).
\btitle{Incorporating retrospective data into an analysis of time to illness}.
\bjournal{Biostatistics}
\bvolume{2}
\bpages{1--12}.
\bid{doi={10.1093/biostatistics/2.1.1}, issn={1468-4357}, pii={2/1/1},
pmid={12933553}}
\end{barticle}
%
\bptok{imsref}%
% NOT OUTPUTED:
% issn = 1468-4357
% number = 1
% fjournal = Biostatistics (Oxford, England)
\endbibitem

%b6 #&#
\bibitem[\protect\citeauthoryear{Cox et~al.}{2007}]{CoxChu2007}
%
\begin{barticle}[mr]
\bauthor{\bsnm{Cox},~\bfnm{Christopher}\binits{C.}},
\bauthor{\bsnm{Chu},~\bfnm{Haitao}\binits{H.}},
\bauthor{\bsnm{Schneider},~\bfnm{Michael~F.}\binits{M.~F.}} \AND
\bauthor{\bsnm{Mu{\~n}oz},~\bfnm{Alvaro}\binits{A.}}
(\byear{2007}).
\btitle{Parametric survival analysis and taxonomy of hazard functions
for the generalized gamma distribution}.
\bjournal{Stat. Med.}
\bvolume{26}
\bpages{4352--4374}.
\bid{doi={10.1002/sim.2836}, issn={0277-6715}, mr={2405358}}
\end{barticle}
%
\bptok{imsref}%
% NOT OUTPUTED:
% issn = 0277-6715
% url = http://dx.doi.org/10.1002/sim.2836
% number = 23
% fjournal = Statistics in Medicine
\endbibitem

%b7 #&#
\bibitem[\protect\citeauthoryear{De~Angelis et~al.}{2009}]{Deangelis2009}
%
\begin{barticle}[mr]
\bauthor{\bsnm{De Angelis},~\bfnm{D.}\binits{D.}},
\bauthor{\bsnm{Sweeting},~\bfnm{M.}\binits{M.}},
\bauthor{\bsnm{Ades},~\bfnm{A.~E.}\binits{A.~E.}},
\bauthor{\bsnm{Hickman},~\bfnm{M.}\binits{M.}},
\bauthor{\bsnm{Hope},~\bfnm{V.}\binits{V.}} \AND
\bauthor{\bsnm{Ramsay},~\bfnm{M.}\binits{M.}}
(\byear{2009}).
\btitle{An evidence synthesis approach to estimating hepatitis {C}
prevalence in {E}ngland and {W}ales}.
\bjournal{Stat. Methods Med. Res.}
\bvolume{18}
\bpages{361--379}.
\bid{doi={10.1177/0962280208094691}, issn={0962-2802}, mr={2750101}}
\end{barticle}
%
\bptok{imsref}%
% NOT OUTPUTED:
% issn = 0962-2802
% url = http://dx.doi.org/10.1177/0962280208094691
% number = 4
% fjournal = Statistical Methods in Medical Research
\endbibitem

%b8 #&#
\bibitem[\protect\citeauthoryear{Farewell and
Prentice}{1977}]{FarewellPrentice1977}
%
\begin{barticle}[author]
\bauthor{\bsnm{Farewell},~\bfnm{V.~T.}\binits{V.~T.}} \AND
\bauthor{\bsnm{Prentice},~\bfnm{R.~L.}\binits{R.~L.}}
(\byear{1977}).
\btitle{A study of distributional shape in life testing}.
\bjournal{Technometrics}
\bvolume{19}
\bpages{69--75}.
\end{barticle}
%
\bptok{imsref}%
\endbibitem

%b9 #&#
\bibitem[\protect\citeauthoryear{Freeman et~al.}{2001}]{FreemanDoreLaw2001}
%
\begin{barticle}[pbm]
\bauthor{\bsnm{Freeman},~\bfnm{A.~J.}\binits{A.~J.}},
\bauthor{\bsnm{Dore},~\bfnm{G.~J.}\binits{G.~J.}},
\bauthor{\bsnm{Law},~\bfnm{M.~G.}\binits{M.~G.}},
\bauthor{\bsnm{Thorpe},~\bfnm{M.}\binits{M.}},
\bauthor{\bsnm{Overbeck},~\bfnm{J.~Von}\binits{J.~V.}},
\bauthor{\bsnm{Lloyd},~\bfnm{A.~R.}\binits{A.~R.}},
\bauthor{\bsnm{Marinos},~\bfnm{G.}\binits{G.}} \AND
\bauthor{\bsnm{Kaldor},~\bfnm{J.~M.}\binits{J.~M.}}
(\byear{2001}).
\btitle{Estimating progression to cirrhosis in chronic hepatitis C
virus infection}.
\bjournal{Hepatology}
\bvolume{34}
\bpages{809--816}.
\bid{doi={10.1053/jhep.2001.27831}, issn={0270-9139},
pii={S0270-9139(01)33319-0}, pmid={11584380}}
\end{barticle}
%
\bptok{imsref}%
% NOT OUTPUTED:
% issn = 0270-9139
% number = 4 Pt 1
% fjournal = Hepatology (Baltimore, Md.)
\endbibitem

%b10 #&#
\bibitem[\protect\citeauthoryear{Fu, Tom and Bird}{2009}]{FuTomBird2009}
%
\begin{barticle}[mr]
\bauthor{\bsnm{Fu},~\bfnm{Bo}\binits{B.}},
\bauthor{\bsnm{Tom},~\bfnm{Brian~D.~M.}\binits{B.~D.~M.}} \AND
\bauthor{\bsnm{Bird},~\bfnm{Sheila~M.}\binits{S.~M.}}
(\byear{2009}).
\btitle{Re-weighted inference about hepatitis {C} virus-infected
communities when analysing diagnosed patients referred to liver clinics}.
\bjournal{Stat. Methods Med. Res.}
\bvolume{18}
\bpages{303--320}.
\bid{doi={10.1177/0962280208094688}, issn={0962-2802}, mr={2750070}}
\end{barticle}
%
\bptok{imsref}%
% NOT OUTPUTED:
% issn = 0962-2802
% url = http://dx.doi.org/10.1177/0962280208094688
% number = 3
% fjournal = Statistical Methods in Medical Research
\endbibitem

%b11 #&#
\bibitem[\protect\citeauthoryear{Fu et~al.}{2007}]{FuTomBird2007}
%
\begin{barticle}[author]
\bauthor{\bsnm{Fu},~\bfnm{B.}\binits{B.}},
\bauthor{\bsnm{Tom},~\bfnm{B.~D.~M.}\binits{B.~D.~M.}},
\bauthor{\bsnm{Delahooke},~\bfnm{T.}\binits{T.}},
\bauthor{\bsnm{Alexander},~\bfnm{G.~J.~M.}\binits{G.~J.~M.}} \AND
\bauthor{\bsnm{Bird},~\bfnm{S.~M.}\binits{S.~M.}}
(\byear{2007}).
\btitle{Event-biased referral can distort estimation of hepatitis {C}
virus progression rate to cirrhosis, and of prognostic influences}.
\bjournal{J. Clin. Epidemiol.}
\bvolume{60}
\bpages{1140--1148}.
\end{barticle}
%
\bptok{imsref}%
\endbibitem

%b12 #&#
\bibitem[\protect\citeauthoryear{Glaser}{1980}]{Glaser1980}
%
\begin{barticle}[mr]
\bauthor{\bsnm{Glaser},~\bfnm{Ronald~E.}\binits{R.~E.}}
(\byear{1980}).
\btitle{Bathtub and related failure rate characterizations}.
\bjournal{J. Amer. Statist. Assoc.}
\bvolume{75}
\bpages{667--672}.
\bid{issn={0162-1459}, mr={0590699}}
\end{barticle}
%
\bptok{imsref}%
% NOT OUTPUTED:
% issn = 0162-1459
% url =
%http://links.jstor.org/sici?sici=0162-1459(198009)75:371<667:BARFRC>2.0.CO;2-W&origin=MSN
% number = 371
% coden = JSTNAL
% fjournal = Journal of the American Statistical Association
\endbibitem

%b13 #&#
\bibitem[\protect\citeauthoryear{Hagan et~al.}{2008}]{Hagan2008}
%
\begin{barticle}[author]
\bauthor{\bsnm{Hagan},~\bfnm{H.}\binits{H.}},
\bauthor{\bsnm{Pouget},~\bfnm{E.~R.}\binits{E.~R.}},
\bauthor{\bsnm{{D}es Jarais},~\bfnm{D.~C.}\binits{D.~C.}} \AND
\bauthor{\bsnm{Lelutiu-Weinberger},~\bfnm{C.}\binits{C.}}
(\byear{2008}).
\btitle{Meta-regression of hepatitis C virus infection in relation to
time since onset of illicit drug injection: The influence of time and place}.
\bjournal{Am. J. Epidemiol.}
\bvolume{168}
\bpages{1099--1109}.
\end{barticle}
%
\bptok{imsref}%
\endbibitem

%b14 #&#
\bibitem[\protect\citeauthoryear{Hardin and Hilbe}{2003}]{HardinHilbe2003}
%
\begin{bbook}[mr]
\bauthor{\bsnm{Hardin},~\bfnm{James~W.}\binits{J.~W.}} \AND
\bauthor{\bsnm{Hilbe},~\bfnm{Joseph~M.}\binits{J.~M.}}
(\byear{2003}).
\btitle{Generalized Estimating Equations}.
\bpublisher{Chapman \& Hall},
\blocation{London}.
\bid{mr={2000388}}
\end{bbook}
%
\bptok{imsref}%
% NOT OUTPUTED:
% isbn = 1-58488-307-3
% fpage = xiv+222
\endbibitem

%b15 #&#
\bibitem[\protect\citeauthoryear{Hutchinson, Bird and
Goldberg}{2005}]{Hutchinson2005}
%
\begin{barticle}[author]
\bauthor{\bsnm{Hutchinson},~\bfnm{S.~J.}\binits{S.~J.}},
\bauthor{\bsnm{Bird},~\bfnm{S.~M.}\binits{S.~M.}} \AND
\bauthor{\bsnm{Goldberg},~\bfnm{D.~J.}\binits{D.~J.}}
(\byear{2005}).
\btitle{Modelling the current and future disease burden of hepatitis
{C} among injecting drug users in Scotland}.
\bjournal{Hepatology}
\bvolume{42}
\bpages{711--723}.
\end{barticle}
%
\bptok{imsref}%
\endbibitem

%b16 #&#
\bibitem[\protect\citeauthoryear{Kalbfleisch and
Prentice}{2002}]{KalbfleischPrentice2002}
%
\begin{bbook}[mr]
\bauthor{\bsnm{Kalbfleisch},~\bfnm{John~D.}\binits{J.~D.}} \AND
\bauthor{\bsnm{Prentice},~\bfnm{Ross~L.}\binits{R.~L.}}
(\byear{2002}).
\btitle{The Statistical Analysis of Failure Time Data},
\bedition{2nd} ed.
\bpublisher{Wiley},
\blocation{New York}.
\bid{doi={10.1002/9781118032985}, mr={1924807}}
\end{bbook}
%
\bptok{imsref}%
% NOT OUTPUTED:
% isbn = 0-471-36357-X
% url = http://dx.doi.org/10.1002/9781118032985
% fpage = xiv+439
\endbibitem

%b17 #&#
\bibitem[\protect\citeauthoryear{Keiding}{2005}]{Keiding2005}
%
\begin{bincollection}[author]
\bauthor{\bsnm{Keiding},~\bfnm{N.}\binits{N.}}
(\byear{2005}).
\btitle{Delayed entry}.
In \bbooktitle{Encyclopedia of Biostatistics},
\bedition{2nd} ed.
(\beditor{\bfnm{P.}\binits{P.}~\bsnm{Armitage}} \AND
\beditor{\bfnm{T.}\binits{T.}~\bsnm{Colton}}, eds.)
\bpages{1404--1409}.
\bpublisher{Wiley},
\blocation{New York}.
\end{bincollection}
%
\bptok{imsref}%
\endbibitem

%b18 #&#
\bibitem[\protect\citeauthoryear{Lawless}{1980}]{Lawless1980}
%
\begin{barticle}[mr]
\bauthor{\bsnm{Lawless},~\bfnm{J.~F.}\binits{J.~F.}}
(\byear{1980}).
\btitle{Inference in the generalized gamma and log gamma distributions}.
\bjournal{Technometrics}
\bvolume{22}
\bpages{409--419}.
\bid{doi={10.2307/1268326}, issn={0040-1706}, mr={0585636}}
\end{barticle}
%
\bptok{imsref}%
% NOT OUTPUTED:
% issn = 0040-1706
% url = http://dx.doi.org/10.2307/1268326
% number = 3
% coden = TCMTA2
% fjournal = Technometrics. A Journal of Statistics for the Physical,
%Chemical and Engineering Sciences
\endbibitem

%b19 #&#
\bibitem[\protect\citeauthoryear{Lawless}{1997}]{Lawless1997}
%
\begin{bincollection}[mr]
\bauthor{\bsnm{Lawless},~\bfnm{J.~F.}\binits{J.~F.}}
(\byear{1997}).
\btitle{Likelihood and pseudo likelihood estimation based on
response-biased observation}.
In \bbooktitle{Selected {P}roceedings of the {S}ymposium on
{E}stimating {F}unctions ({A}thens, {GA}, 1996)}
(\beditor{\bfnm{V.~B.}\binits{V.~B.}~\bsnm{Ishwar}},
\beditor{\bfnm{V.~P.}\binits{V.~P.}~\bsnm{Godambe}} \AND
\beditor{\bfnm{R.~L.}\binits{R.~L.}~\bsnm{Taylor}}, eds.).
\bseries{Institute of Mathematical Statistics Lecture Notes---Monograph Series}
\bvolume{32}
\bpages{43--55}.
\bpublisher{IMS},
\blocation{Hayward, CA}.
\bid{doi={10.1214/lnms/1215455037}, mr={1837796}}
\end{bincollection}
%
\bptok{imsref}%
% NOT OUTPUTED:
% url = http://dx.doi.org/10.1214/lnms/1215455037
\endbibitem

%b20 #&#
\bibitem[\protect\citeauthoryear{Prentice}{1974}]{Prentice1974}
%
\begin{barticle}[mr]
\bauthor{\bsnm{Prentice},~\bfnm{R.~L.}\binits{R.~L.}}
(\byear{1974}).
\btitle{A~log gamma model and its maximum likelihood estimation}.
\bjournal{Biometrika}
\bvolume{61}
\bpages{539--544}.
\bid{issn={0006-3444}, mr={0378212}}
\end{barticle}
%
\bptok{imsref}%
% NOT OUTPUTED:
% issn = 0006-3444
% fjournal = Biometrika
\endbibitem

%b21 #&#
\bibitem[\protect\citeauthoryear{Qin and Shen}{2010}]{QinShen2010}
%
\begin{barticle}[mr]
\bauthor{\bsnm{Qin},~\bfnm{Jing}\binits{J.}} \AND
\bauthor{\bsnm{Shen},~\bfnm{Yu}\binits{Y.}}
(\byear{2010}).
\btitle{Statistical methods for analyzing right-censored length-biased
data under {C}ox model}.
\bjournal{Biometrics}
\bvolume{66}
\bpages{382--392}.
\bid{doi={10.1111/j.1541-0420.2009.01287.x}, issn={0006-341X}, mr={2758818}}
\end{barticle}
%
\bptok{imsref}%
% NOT OUTPUTED:
% issn = 0006-341X
% url = http://dx.doi.org/10.1111/j.1541-0420.2009.01287.x
% number = 2
% fjournal = Biometrics. Journal of the International Biometric Society
\endbibitem

%b22 #&#
\bibitem[\protect\citeauthoryear{Stacy}{1962}]{Stacy1962}
%
\begin{barticle}[mr]
\bauthor{\bsnm{Stacy},~\bfnm{E.~W.}\binits{E.~W.}}
(\byear{1962}).
\btitle{A generalization of the gamma distribution}.
\bjournal{Ann. Math. Statist.}
\bvolume{33}
\bpages{1187--1192}.
\bid{issn={0003-4851}, mr={0143277}}
\end{barticle}
%
\bptok{imsref}%
% NOT OUTPUTED:
% issn = 0003-4851
% fjournal = Annals of Mathematical Statistics
\endbibitem

%b23 #&#
\bibitem[\protect\citeauthoryear{Stacy and Mihram}{1965}]{StacyMihram1965}
%
\begin{barticle}[mr]
\bauthor{\bsnm{Stacy},~\bfnm{E.~W.}\binits{E.~W.}} \AND
\bauthor{\bsnm{Mihram},~\bfnm{G.~A.}\binits{G.~A.}}
(\byear{1965}).
\btitle{Parameter estimation for a generalized gamma distribution}.
\bjournal{Technometrics}
\bvolume{7}
\bpages{349--358}.
\bid{issn={0040-1706}, mr={0192586}}
\end{barticle}
%
\bptok{imsref}%
% NOT OUTPUTED:
% issn = 0040-1706
% fjournal = Technometrics. A Journal of Statistics for the Physical,
%Chemical and Engineering Sciences
\endbibitem

%b24 #&#
\bibitem[\protect\citeauthoryear{Struthers and
Farewell}{1989}]{StruthersFarewell1989}
%
\begin{barticle}[author]
\bauthor{\bsnm{Struthers},~\bfnm{C.~A.}\binits{C.~A.}} \AND
\bauthor{\bsnm{Farewell},~\bfnm{V.~T.}\binits{V.~T.}}
(\byear{1989}).
\btitle{A mixture model for time to {AIDS} data with left truncation
and an uncertain origin}.
\bjournal{Biometrika}
\bvolume{76}
\bpages{814--817}.
\end{barticle}
%
\bptok{imsref}%
\endbibitem

%b25 #&#
\bibitem[\protect\citeauthoryear{Tom, Farewell and Bird}{2014}]{supp}
\begin{bmisc}[author]
\bauthor{\bsnm{Tom},~\bfnm{B. D. M.}\binits{B. D. M.}},
\bauthor{\bsnm{Farewell},~\bfnm{V. T.}\binits{V. T.}} \AND
\bauthor{\bsnm{Bird},~\bfnm{S. M.}\binits{S. M.}}
(\byear{2014}).
\bhowpublished{Supplement to ``Maximum likelihood and pseudo score approaches for parametric
time-to-event analysis with informative entry times.''
DOI:\doiurl{10.1214/14-AOAS725SUPP}}.
\bptok{imsref}%
\end{bmisc}
\endbibitem

%b26 #&#
\bibitem[\protect\citeauthoryear{Tsai}{2009}]{Tsai2009}
%
\begin{barticle}[mr]
\bauthor{\bsnm{Tsai},~\bfnm{Wei~Yann}\binits{W.~Y.}}
(\byear{2009}).
\btitle{Pseudo-partial likelihood for proportional hazards models with
biased-sampling data}.
\bjournal{Biometrika}
\bvolume{96}
\bpages{601--615}.
\bid{doi={10.1093/biomet/asp026}, issn={0006-3444}, mr={2538760}}
\end{barticle}
%
\bptok{imsref}%
% NOT OUTPUTED:
% issn = 0006-3444
% url = http://dx.doi.org/10.1093/biomet/asp026
% number = 3
% coden = BIOKAX
% fjournal = Biometrika
\endbibitem

%b27 #&#
\bibitem[\protect\citeauthoryear{Wang}{2005}]{Wang2005}
%
\begin{bincollection}[author]
\bauthor{\bsnm{Wang},~\bfnm{M.-C.}\binits{M.-C.}}
(\byear{2005}).
\btitle{Length bias}.
In \bbooktitle{Encyclopedia of Biostatistics},
\bedition{2nd} ed.
(\beditor{\bfnm{P.}\binits{P.}~\bsnm{Armitage}} \AND
\beditor{\bfnm{T.}\binits{T.}~\bsnm{Colton}}, eds.)
\bpages{2756--2759}.
\bpublisher{Wiley},
\blocation{New York}.
\end{bincollection}
%
\bptok{imsref}%
\endbibitem

%b28 #&#
\bibitem[\protect\citeauthoryear{Wang, Brookmeyer and
Jewell}{1993}]{WangBrookmeyerJewell1993}
%
\begin{barticle}[mr]
\bauthor{\bsnm{Wang},~\bfnm{Mei-Cheng}\binits{M.-C.}},
\bauthor{\bsnm{Brookmeyer},~\bfnm{Ron}\binits{R.}} \AND
\bauthor{\bsnm{Jewell},~\bfnm{Nicholas~P.}\binits{N.~P.}}
(\byear{1993}).
\btitle{Statistical models for prevalent cohort data}.
\bjournal{Biometrics}
\bvolume{49}
\bpages{1--11}.
\bid{doi={10.2307/2532597}, issn={0006-341X}, mr={1221402}}
\end{barticle}
%
\bptok{imsref}%
% NOT OUTPUTED:
% issn = 0006-341X
% url = http://dx.doi.org/10.2307/2532597
% number = 1
% coden = BIOMB6
% fjournal = Biometrics. Journal of the Biometric Society
\endbibitem

\end{thebibliography}
\end{document}